\documentclass{ws-procs9x6-cpt16}

\begin{document}

\def\etal {{\it et al.}}

\newcommand{\refeq}[1]{(\ref{#1})}

\def\al{\alpha}
\def\be{\beta}
\def\ga{\gamma}
\def\de{\delta}
\def\ep{\epsilon}
\def\ve{\varepsilon}
\def\ze{\zeta}
\def\et{\eta}
\def\th{\theta}
\def\vt{\vartheta}
\def\io{\iota}
\def\ka{\kappa}
\def\la{\lambda}
\def\vpi{\varpi}
\def\rh{\rho}
\def\vr{\varrho}
\def\si{\sigma}
\def\vs{\varsigma}
\def\ta{\tau}
\def\up{\upsilon}
\def\ph{\phi}
\def\vp{\varphi}
\def\ch{\chi}
\def\ps{\psi}
\def\om{\omega}
\def\Ga{\Gamma}
\def\De{\Delta}
\def\Th{\Theta}
\def\La{\Lambda}
\def\Si{\Sigma}
\def\Up{\Upsilon}
\def\Ph{\Phi}
\def\Ps{\Psi}
\def\Om{\Omega}
\def\cA{{\cal A}}
\def\cB{{\cal B}}
\def\cC{{\cal C}}
\def\cE{{\cal E}}
\def\cG{{\cal G}}
\def\cl{{\cal L}}
\def\cL{{\cal L}}
\def\cO{{\cal O}}
\def\cP{{\cal P}}
\def\cR{{\cal R}}
\def\cV{{\cal V}}
\def\ni{\noindent}
\def\fr#1#2{{{#1} \over {#2}}}
\def\half{{\textstyle{1\over 2}}}
\def\quar{{\textstyle{1\over 4}}}
\def\frac#1#2{{\textstyle{{#1}\over {#2}}}}

\def\lsim{\mathrel{\rlap{\lower4pt\hbox{\hskip1pt$\sim$}}
    \raise1pt\hbox{$<$}}}
\def\gsim{\mathrel{\rlap{\lower4pt\hbox{\hskip1pt$\sim$}}
    \raise1pt\hbox{$>$}}}
\def\Re{\hbox{Re}\,}
\def\Im{\hbox{Im}\,}
\def\prt{\partial}

\def\etal{{\it et al.}}
\def\pt#1{\phantom{#1}}

\newcommand{\beq}{\begin{equation}}
\newcommand{\eeq}{\end{equation}}
\newcommand{\bea}{\begin{eqnarray}}
\newcommand{\eea}{\end{eqnarray}}
\newcommand{\bit}{\begin{itemize}}
\newcommand{\eit}{\end{itemize}}
\newcommand{\rf}[1]{(\ref{#1})}
\newcommand{\nn}{\nonumber}

\def\kr{{(k^{(4)})}{}}
\def\kdr{{(k^{(5)})}{}}
\def\kddr{{(k^{(6)}_1)}{}}
\def\krr{{(k^{(6)}_2)}{}}
\def\krrr{{(k^{(8)})}{}}

\def\krnp{{k^{(4)}}}
\def\kdrnp{{k^{(5)}}}
\def\kddrnp{{k^{(6)}_1}}
\def\krrnp{{k^{(6)}_2}}
\def\krrrnp{{k^{(8)}}}

\def\kb{\overline{k}{}}
\def\kt{\widetilde{k}{}}

\def\kbr{{(\kb^{(4)})}}
\def\kbdr{{(\kb^{(5)})}}
\def\kbddr{{(\kb^{(6)}_1)}}
\def\kbrr{{(\kb^{(6)}_2)}}
\def\kbrrr{ {(\kb^{(8)})}  }
\def\kbef{ {(\kb_{\rm eff})} }

\def\ktr{{(\kt^{(4)})}{}}
\def\ktdr{{(\kt^{(5)})}{}}
\def\ktddr{{(\kt^{(6)}_1)}{}}
\def\ktrr{{(\kt^{(6)}_2)}{}}

\def\tb{\overline{t}}
\def\sb{\overline{s}}
\def\ub{\overline{u}}

\def\kl{{\ka\la}}
\def\mn{{\mu\nu}}
\def\ab{{\al\be}}
\def\abl{{\al\be\ldots}}
\def\abgd{{\al\be\ga\de}}
\def\abkl{{\al\be\ka\la}}
\def\gdmn{{\ga\de\mu\nu}}
\def\klmn{{\ka\la\mu\nu}}
\def\ezet{{\ep\ze\et\th}}
\def\ol#1{\overline{#1}}

\def\cG{{G}}

\def\mbf#1{\boldsymbol #1}

\title{Gravity Sector of the SME}

\author{Q.G.\ Bailey}

\address{Physics Department,
Embry-Riddle Aeronautical University\\
3700 Willow Creek Road, 
Prescott, AZ 86301, USA\\}

\begin{abstract}
In this talk, 
the gravity sector of the effective field theory description of local Lorentz violation is discussed,
including minimal and nonminimal curvature couplings.  
Also,
recent experimental and observational analyses including solar-system ephemeris and short-range gravity tests
are reviewed.
\end{abstract}

\bodymatter

\section{Introduction} 

A comprehensive and highly successful description of nature 
is provided by General Relativity and the Standard Model of particle physics. 
However, 
at the Planck energy scale it is widely believed that an underlying unified description 
exists that contains both theories as limiting cases.
A compelling and predictive unified theory remains largely unknown to date
and experimental clues about such an underlying theory remain scarce since
direct measurements at the Planck scale are infeasible.

The idea to study suppressed effects that might arise from an underlying unified theory 
is a promising alternative approach. 
In particular, 
minuscule violations of local Lorentz symmetry are an intriguing class of signals 
that are potentially detectable in modern high-precision experiments.\cite{strings, t14}
The Standard-Model Extension (SME) is a comprehensive effective field theory framework 
that describes the observable signals of Lorentz violation.\cite{sme}  
The indexed coefficients for Lorentz violation control the degree of Lorentz breaking
for each type of matter or field in this framework and they represent the experimentally sought quantities.\cite{datatables}

While much work in the last two decades has involved the flat-spacetime limit of the SME, 
recently the activity in the gravitational sector has increased.
The gravitational sector of the SME includes both pure-gravity couplings and matter-gravity couplings.\cite{bk06,kt0911}
In this talk, 
we focus on the pure-gravity sector discussing the basic theory and recent analyses.

\section{Framework}

In the effective field theory description of local Lorentz violation in the pure-gravity sector,
the Lagrange density includes the Einstein-Hilbert lagrangian $\cL_{\rm EH}$ 
and the matter sector lagrangian $\cL_{\rm M}$.
The Lorentz-violating term $\cL_{\rm LV}$ is constructed using coefficient fields
contracted with ``operators" built from curvature tensors and covariant derivatives 
with increasing mass dimension.
Including the mass dimension $4$ through $6$ operators the lagrangian $\cL_{\rm LV}$ is given by
\bea
\cL_{\rm LV} &=& \frac {\sqrt{-g}}{16 \pi G} 
[ \kr_\abgd R^\abgd + \kdr_{\al\be\ga\de\ka} D^\ka R^\abgd
\nonumber\\
&&
+\kddr_{\al\be\ga\de\ka\la} D^{(\ka} D^{\la)} R^\abgd
+\krr_{\abgd\klmn} R^\abgd R^\klmn ].
\eea
The minimal SME is contained in the $d=4$ $\kr_\abgd$ case, 
which can be split into a total trace $u$, 
a trace $s_\mn$, 
and a traceless piece $t_\klmn$.\cite{bonder}
The mass dimension $5$ term involving the coefficients $\kdr_{\abgd\ka}$ breaks CPT symmetry, 
and the mass dimension $6$ terms are controlled by the $\kddrnp$ and $\krrnp$ coefficients.
This lagrangian is supplemented by a term $\cl_k$ that contains contributions from the dynamics of the coefficient fields
$\krnp$, 
$\kdrnp$, 
$\kddrnp$, 
and $\krrnp$.

To find effective Einstein equations and equations of motion for matter particles, 
phenomenology can proceed by assuming spontaneous Lorentz-symmetry breaking.
In this scenario, 
the coefficient fields acquire nonzero vacuum expectation values
through a dynamical process.
For instance, 
considering the $s_\mn$ coefficients, 
the vacuum expectation values are denoted $\sb_\mn$.
In the analysis so far in the linearized gravity limit, 
it has been shown that the fluctuations around the vacuum value can be ``decoupled" from the gravitational 
fluctuations $h_\mn$ under mild assumptions, 
so that the effective linearized field equations depend only on the vacuum values $\sb_\mn$.
Once the effective linearized field equations are obtained, 
the post-newtonian metric can be calculated up to $PNO(3)$\cite{bk06,bkx15} 
and the effects on propagation can be studied.\cite{kt15, km16}

\section{Experiment and observation}

In the minimal SME limit of the gravity sector,
the nine independent coefficients in the traceless $\sb_\mn$ control the 
dominant effects in weak-field gravity.
These can be measured in a variety of post-newtonian tests in laboratories, 
the solar system, 
and beyond.
For example, 
stringent constraints on seven $\sb_\mn$ coefficients have been obtained using data from
atom-interferometric gravimeters by searching for sidereal day and
annual variations in the free-fall acceleration of cesium atoms.\cite{atom} 

Some of the main observable effects for orbits include additional secular changes
in keplerian orbital elements.
The changes in the orbital elements depend on different combinations 
of the coefficients $\sb_\mn$ for each orbit due to orientation dependence, 
which indicates the breaking of rotational symmetry. 
In particular, 
precise measurements and modeling of the secular changes of the perihelia and longitude 
of the node for six planets have been used to improve constraints on eight of the
$\sb_\mn$ coefficients by using an analysis of post-fit residuals.\cite{hees15}
These limits reach the $10^{-8}$ level on the $\sb_{TJ}$ coefficients 
and $10^{-11}$ on the $\sb_{JK}$ coefficients.
Analyses of binary pulsar orbits have also been used to place competitive limits 
on these coefficients.\cite{shao14}

While much analysis has used post-fit residuals,
including a 2007 analysis of lunar laser ranging,\cite{battat07}
a more rigorous approach should include the SME equations of motion directly in the modeling code.
This challenging task has been achieved recently, 
where the SME coefficients $\sb_\mn$ are included as fit parameters in the analysis of lunar laser ranging data 
spanning over $40$ years.\cite{bourgoin16}
This has resulted in significant improvement of the best ``laboratory" limits including parts in $10^{12}$
on some components of $\sb_{JK}$, 
rivaled only by analysis of cosmic rays.\cite{kt15}

For the mass dimension $6$ coefficients $\kddrnp$ and $\krrnp$, 
among the best tests are short-range gravity experiments, 
where the gravitational force between two masses is precisely studied 
at the millimeter level and below.
To calculate the observable effects of Lorentz violation in this context, 
a description at the level of the modified newtonian potential can be used.\cite{bkx15}
For a point mass $M$, 
the potential is given by 
\beq
U = \fr {G M}{r} \left(1+ \fr {{\overline k} ({\hat r})}{r^2} \right), 
\label{pot}
\eeq
where the anisotropic quantity ${\overline k} ({\hat r})$ is given by
\beq
{\overline k} ({\hat r}) = \frac 32 \kbef_{jkjk}-9\kbef_{jkll} {\hat r}^j {\hat r}^k
+\frac {15}{2} \kbef_{jklm} {\hat r}^j {\hat r}^k {\hat r}^l {\hat r}^m.
\eeq
The effective coefficients $\kbef_{jklm}$ are linear combinations of $\kbddr$ and $\kbrr$ 
and the result is valid for $\vec r \neq 0$.\cite{bkx15, shao16}

There are in principle $14$ observable {\it a priori} independent coefficients in Eq.\ \rf{pot}
and any single experiment is sensitive to eight combinations via the sidereal day time dependence
from the Earth's rotation.
In particular, 
when one expresses the laboratory frame coefficients $\kbef_{jklm}$ in terms of 
the standard Sun-centered celestial equatorial frame coefficients $\kbef_{JKLM}$, 
signals up to the fourth harmonic in the Earth's sidereal frequency appear, 
offering a striking signature for Lorentz violation in short-range gravity tests.
Analyses of short-range gravity tests searching for $\kbef$ coefficients have been performed
by the IU experiment in Ref.\ \refcite{kl15}
and 
the HUST collaboration in Ref.\ \refcite{hust15}.
Due to the differing locations of these experiments, 
the $14$ observable coefficients in $\kbef_{JKLM}$ can be disentangled using data 
from both experiments.
Such a combined analysis was performed recently in Ref.\ \refcite{shao16}, 
with limits at the $10^{-9}$ m$^2$ level on these coefficients.

\section*{Acknowledgments}

This work was supported in part by the National Science Foundation
under grant number PHY-1402890.


\begin{thebibliography}{xx}

\bibitem{strings}
V.A.\ Kosteleck\'y and S.\ Samuel,
Phys.\ Rev.\ D {\bf 39}, 683 (1989);
V.A.\ Kosteleck\'y and R.\ Potting, 
Phys.\ Rev.\ D {\bf 51}, 3923 (1995).

\bibitem{t14}
For reviews, see 
J.\ Tasson, Rept.\ Prog.\ Phys.\ {\bf 77}, 062901 (2014);
C.M.\ Will, Living Rev.\ Relativ.\ {\bf 17}, 4 (2014).

\bibitem{sme}
D.\ Colladay and V.A.\ Kosteleck\'y,
Phys.\ Rev.\ D {\bf 55}, 6760 (1997);
Phys.\ Rev.\ D {\bf 58}, 116002 (1998);
V.A.\ Kosteleck\'y, Phys.\ Rev.\ D {\bf 69}, 105009 (2004).

\bibitem{datatables}
{\it Data Tables for Lorentz and CPT Violation,}
V.A.\ Kosteleck\'y and N.\ Russell,
2016 edition,
arXiv:0801.0287v9.

\bibitem{bk06}
Q.G.\ Bailey and V.A.\ Kosteleck\'y,
Phys.\ Rev.\ D {\bf 74}, 045001 (2006).

\bibitem{kt0911} 
V.A.\ Kosteleck\'y and J.D.\ Tasson,
Phys.\ Rev.\ Lett.\ {\bf 102}, 010402 (2009);
Phys.\ Rev.\ D {\bf 83}, 016013 (2011);
J.\ Tasson, these proceedings.

\bibitem{bonder}
Y.\ Bonder, 
Phys.\ Rev.\ D {\bf 91}, 125002 (2015); 
these proceedings

\bibitem{bkx15}
Q.G.\ Bailey, V.A.\ Kosteleck\'y, and R.\ Xu, 
Phys.\ Rev.\ D {\bf 91}, 022006 (2015).

\bibitem{kt15} 
V.A.\ Kosteleck\'y and J.D.\ Tasson,
Phys.\ Lett.\ B {\bf 749}, 551  (2015).

\bibitem{km16}
V.A.\ Kosteleck\'y and M.\ Mewes,
Phys.\ Lett.\ B {\bf 757}, 510 (2016).

\bibitem{atom}
H.\ M\"uller \etal,
Phys.\ Rev.\ Lett.\ {\bf 100}, 031101 (2008);
K.-Y.\ Chung \etal,
Phys.\ Rev.\ D {\bf 80}, 016002 (2009).

\bibitem{hees15}
A.\ Hees \etal, 
Phys.\ Rev.\ D {\bf 92}, 064049 (2015).

\bibitem{shao14}
L.\ Shao,
Phys.\ Rev.\ D {\bf 90}, 122009 (2014);
Phys.\ Rev.\ Lett.\ {\bf 112}, 111103 (2014).

\bibitem{battat07}
J.B.R.\ Battat, J.F.\ Chandler, and C.W.\ Stubbs,
Phys.\ Rev.\ Lett.\ {\bf 99}, 241103 (2007).

\bibitem{bourgoin16}
A.\ Bourgoin \etal,
arXiv:1607.00294.

\bibitem{shao16}
C.G.\ Shao \etal, 
arXiv:1607.06095.

\bibitem{kl15}
J.C.\ Long and V.A.\ Kosteleck\'y,
Phys.\ Rev.\ D {\bf 91}, 092003 (2015).

\bibitem{hust15}
C.G.\ Shao \etal, 
Phys.\ Rev.\ D {\bf 91}, 102007 (2015).

\end{thebibliography}
\end{document}